%% file: main.tex
\documentclass[a4paper,11pt]{article}
\usepackage{pos}
\usepackage{gensymb}

\title{Prototype Schwarzschild-Couder Telescope for the Cherenkov Telescope Array: Commissioning the Optical System}
 \ShortTitle{pSCT Optics System}

\author*[a]{Deivid Ribeiro}

\affiliation [a]{Department of Physics,
 \\ Columbia University, New York, NY 10027, USA}

\forColl{CTA SCT} % W/O "Consortium"

\emailAdd{dr2792@columbia.edu}

\abstract{A prototype Schwarzschild-Couder Telescope (pSCT) has been constructed at the Fred Lawrence Whipple Observatory as a candidate for the medium-sized telescopes of the Cherenkov Telescope Array Observatory (CTAO). CTAO is currently entering early construction phase of the project and once completed it will vastly improve very high energy gamma-ray detection component in multi-wavelength and multi-messenger observations due to significantly improved sensitivity, angular resolution and field of view comparing to the current generation of the ground-based gamma-ray observatories H.E.S.S., MAGIC and VERITAS.  The pSCT uses a dual aspheric mirror design with a $9.7$ m primary mirror and $5.4$ m secondary mirror, both of which are segmented. The Schwarzschild-Couder (SC) optical system (OS) selected for the prototype telescope achieves wide field of view of $8$ degrees and simultaneously reduces the focal plane plate scale allowing an unprecedented compact ($0.78$m diameter) implementation of the high-resolution camera ($6$mm/ $0.067$deg per imaging pixel with $11,328$ pixels) based on the silicon photo-multipliers (SiPMs). The OS of the telescope is designed to eliminate spherical and comatic aberrations and minimize astigmatism to radically improve off-axis imaging and consequently angular resolution across all the field of view with respect to the conventional single-mirror telescopes. Fast and high imaging resolution OS of the pSCT comes with the challenging submillimeter-precision custom alignment system, which was successfully demonstrated with an on-axis point spread function (PSF) of $2.9$ arcmin prior to the first-light detection of the Crab Nebula in 2020. Ongoing commissioning activities aim to meet the on-axis PSF design goal of $2.6$ arcmin, verify the off-axis performance of the pSCT OS, and develop techniques to maintain alignment stability over telescope structural deformations from pointing and temperature variations. In this contribution, we report on the commissioning status, the optical alignment procedures adopted for segmented OS, and alignment progress to verify and validate design requirements.}

\FullConference{37$^{\rm{th}}$ International Cosmic Ray Conference (ICRC 2021)\\
		July 12th -- 23rd, 2021\\
		Online -- Berlin, Germany}

\begin{document}
\maketitle

%%%%%%%%%%%%%%%%%%%%%%%%%%%%%%%%%%%%%%%%%%%%%%%%%
% Introduction
%%%%%%%%%%%%%%%%%%%%%%%%%%%%%%%%%%%%%%%%%%%%%%%%%
\section{Introduction}
Recent advances in the observational capabilities of gravitational waves and detectors of very-high-energy (VHE) neutrinos and gamma-rays have expanded the field of multi-messenger astrophysics. To further explore the new sources found by these detectors, which encompass high-energy, transient, point-like or extended sources, observational synergies require improvements to the VHE detectors for photons with energies greater than 100 GeV. 

The Earth's atmosphere interacts with VHE primary photons, creating cascades of secondary particles that emit Cherenkov radiation broadly distributed between $250$ nm and $800$ nm, peaking around $350-400$ nm. The arrays of imaging atmospheric Cherenkov telescopes (IACTs) indirectly observe VHE photons by stereoscopic imaging of the shower cascades and reconstructing their energy and direction from the ground. Current-generation of small arrays of IACTs use prime-focus OS based on the Davies-Cotton (DC) or parabolic design, which consists of a single segmented mirror with large aperture (>10 m) and $f/ \#$ ratio in the range of $1.2-1.4$. As an alternative, the design described in this proceeding, the fast Schwarzschild-Couder OS ($f / 0.578$) uses two segmented aplanatic mirrors to achieve the best possible imaging resolution in the CTA core energy range (0.1 TeV - 10 TeV). These optical improvements enable a high-resolution SiPM camera (see \cite{leslie}), but require sub-mm and sub-mrad alignment precision. 

The initial alignment of the segmented primary (M1) and secondary (M2) mirrors based on the mirror panel edge sensors (MPESs), which reduced relative positional accuracy of the mirror segments from the telescope structure manufacturing errors of $\sim10$ mm to about $0.3$ mm is described in ~\cite{Adams2020, Adams2020a}. These proceedings review the ongoing development of the optical alignment methods utilizing de-focused images of bright stars to achieve relative positional accuracy of M1 and M2 mirror segments relevant to the point spread function (PSF) degrees of freedom (d.o.f.) of the order of $30$ microns or better. A particular attention is given to the alignment of the M2 inner ring of mirror segments, which is the key to achieving design requirement off-axis performance. This report is structured as follows: section \ref{sec:optics_overview} reviews the overall optical design and subsystems; section \ref{sec:alignment_overview} reviews the optical alignment procedures; section \ref{sec:s1_alignment} reports the commissioning of an alignment method for panels of the M2 inner ring; section \ref{sec:elevation_psf} reports on the ability to maintain the alignment with changes in elevation; and section \ref{sec:summary} summarizes our work.

% , composed of 25 modules (1600 pixels) and has a 2.68\textdegree field of view in it's current state. The SCT was able to detect the Crab Nebula in 2020 following a 20 hr campaign \cite{brent}. An upgrade to the pSCT is underway which is expected to be complete by the end of 2022. The upgrade will include improvements to the camera consisting of an update of the camera module front end electronics (FEE), the camera backend electronics, and the module SiPMs as well as fully populating the focal plane (increasing the number of pixels to 11,328 and the field of view to 8\textdegree)\cite{leslie}.

% SCT is a candidate for MST

% is better than conventional design because it eliminates comatic and aspherical aberrations

% goal: develop methods to maintain PSF (~3) across wide range of operating conditions. Previous work did P1, P2, S2. This update show further alignment + development of corrections based on pointing
% conclusion: we did it, here's how well we're doing and what remains: off-axis alignment, GAS, single panel PSF and P1 alignment -> further work on P2. 

%%%%%%%%%%%%%%%%%%%%%%%%%%%%%%%%%%%%%%%%%%%%%%%%%
% Review of optics - whole system
% deivid
%%%%%%%%%%%%%%%%%%%%%%%%%%%%%%%%%%%%%%%%%%%%%%%%%
\section{Overview of The pSCT Optical System} \label{sec:optics_overview}
The SCT OS is designed to fully correct for spherical and comatic aberrations in the full $8$ degrees field of view (FoV). Astigmatism dominates PSF at the edge of the FoV and it is minimized by implementing a curved focal plane. The choice of a two-mirror design with a de-magnifying secondary mirror reduces the focal plane (FP) plate scale to provide compatibility with the SiPM detectors \cite{Vassiliev07}. The $9.7$m diameter primary mirror (M1) is segmented into $48$ mirror panels, split between two rings: an inner ring (P1) of $16$ panels and an outer ring (P2) of $32$ panels. Similarly, the $5.4$m-diameter secondary mirror (M2) is segmented into $24$ mirror panels, split between two rings: an inner ring (S1) of $8$ panels and an outer ring (S2) of $16$ panels. The focal length of the fast SC OS is $5.586$m. To achieve the point spread function (PSF) of the OS in the FoV compatible with the SiPM pixel size ($6$mm) the sub-mm and sub-mrad alignment is required \cite{2015arXiv150902345O}.

To achieve the alignment goals of the pSCT, a central computer monitors and controls the two major subsystems: the \emph{global} alignment system (GAS) that measures the relative positions of the main optical elements such as M1, M2 and FP; and the \emph{panel-to-panel} alignment system (P2PAS) that measures the relative misalignment between neighbouring panels using mirror panel edge sensors (MPESs). The motion of each panel is controlled by the six actuators assembled into a Stewart Platform \cite{Sreenivasan94}. The controller board (CB) and micro-computer installed on each panel controls all six d.o.f. of each panel with the precision of $3$ microns per step. The panel CBs and central computer communicate with software built using the OPC-UA protocol \cite{mahnke2009opc}. Each CB can receive commands in parallel, enabling fast, system-wide motion of all panels.
% hardware components MPES, ACT, CB, GAS
% Software, OPCUA+SExtractor

%%%%%%%%%%%%%%%%%%%%%%%%%%%%%%%%%%%%%%%%%%%%%%%%%
% Review of optics - Pattern alignment
% Ruo
%%%%%%%%%%%%%%%%%%%%%%%%%%%%%%%%%%%%%%%%%%%%%%%%%
\section{Overview of Optical Alignment Procedure} \label{sec:alignment_overview}
\input{ICRC2021 template/optical_alignment_overview}

% - Refer to Qi and Ruo SPIE proceedings for more detailed focal plane images process
% - Panels identified on image and response matrices measured
% - 1st order corrections measured

\begin{figure}[t!]
    \centering
    \includegraphics[width=\textwidth]{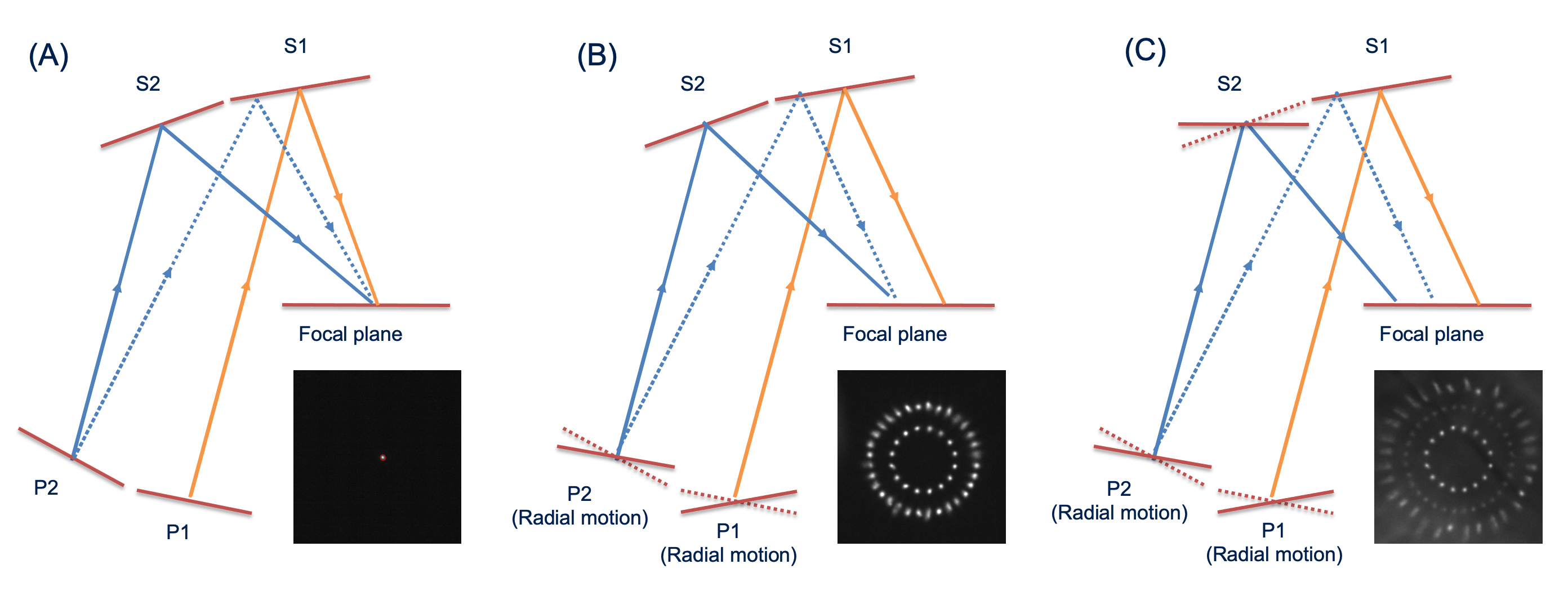}
    \caption{A general overview of the alignment process using a de-focused star projected on the focal plane. The diagram perspective shows a cross-section of primary mirror panels P1 and P2; secondary mirror panels S1 and S2; and the focal plane. The telescope is pointing toward a star along optical axis (upward in diagram). A) The aligned panels project a single focus point of the star. B) Primary panels P1 and P2 are rotated radially outward to de-focus the star into two concentric rings, projecting the individual panels from each P1 and P2 sectors reflecting off the aligned secondary panels.  C) Secondary panels S2 are rotated radially outward to defocus the star even further, separating the projection of P2 panels from their S1 and S2 counterparts.}
    \label{fig:optical_alignment_overview}
\end{figure}

%%%%%%%%%%%%%%%%%%%%%%%%%%%%%%%%%%%%%%%%%%%%%%%%%
% Review of S1 alignment
% Qi (Jun 14 @ 1am ET)
%%%%%%%%%%%%%%%%%%%%%%%%%%%%%%%%%%%%%%%%%%%%%%%%%
\section{Alignment of the S1 Panels}\label{sec:s1_alignment}
% Define problem - all three ring patterns depend on well aligned S1, since they are P1 to S1, P1 to S1, and  P2 to S2
% Define procedure - P1 corrugation, explored -1, -2, -3, and -4  mm rotations
% Effect of Beams - off axis measurements
% next steps

Utilizing the three-ring de-focused on-axis images of a star the P1, P2, and S2 panels can be aligned precisely to the focus while the S1 panels remain fixed. Any mis-alignment in S1 ring therefore will be corrected on-axis by the accurate "mis-alignment" of the P1, P2 and S2 panels. This degeneracy in the on-axis alignment is not realized for off-axis imaging and, therefore, the precise optical alignment of S1 ring is the key to off-axis optical imaging. 

The alignment of S1 panels, however, requires a more complicated procedure, which was successfully developed and is described in this section. This procedure, together with the three-ring configuration, makes for a nearly complete set of methods of optical alignment of the pSCT using de-focused images of stars. 

\begin{figure}[t!]
    \centering
    \includegraphics[width=\textwidth]{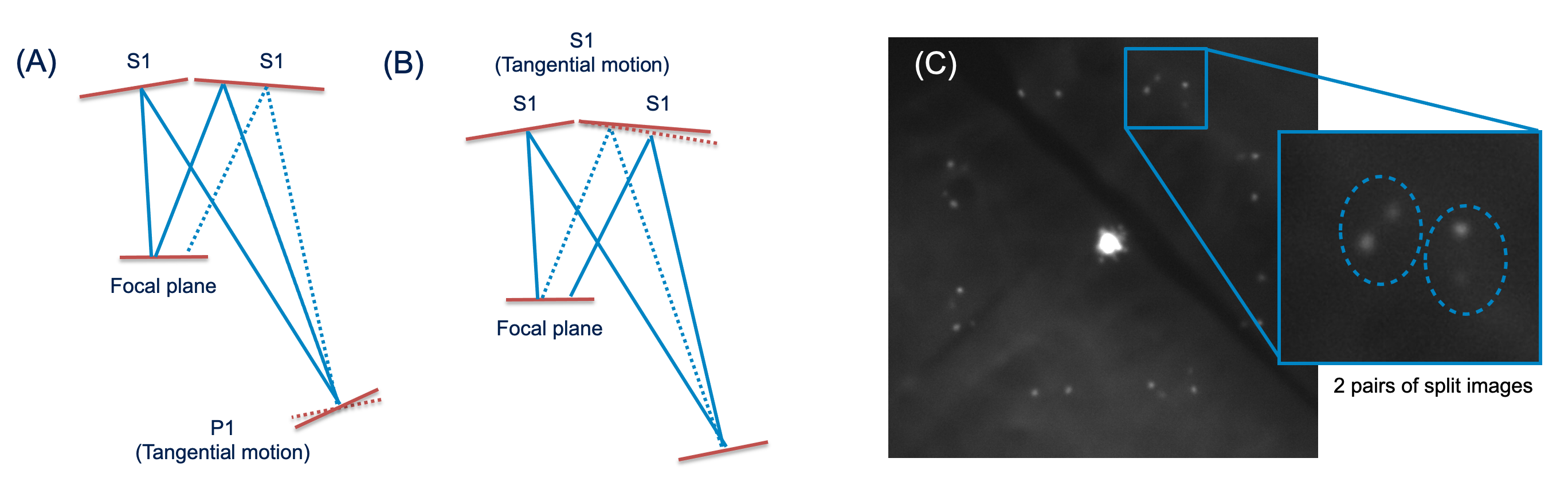}
    \caption{A general overview of the S1 alignment process. A) Assuming that S1 panels are misaligned, P1 panels are rotated tangentially to create a second set of centroids on the focal plane. B) S1 panels are rotated tangentially to remove the new centroids, therefore aligning the S1 panels. C) Focal plane image of the projected centroids. Inset: 2 pairs of split images visible by the misalignment of S1 panels, to be merged and corrected.}
    \label{fig:S2_alignment_overview}
\end{figure}

Nominally, the light from an on-axis star reflected from a P1 panel will be reflected from a single S1 panel before arriving at the focal plane. As illustrated by Figure~\ref{fig:S2_alignment_overview} (A), a tangential rotation (about the radial axis) of a P1 panel will change the optical path so that the reflected light from this panel will be reflected from two neighboring S1 panels. If the neighboring S1 panels are not properly aligned relative to one another, the light from the P1 panel will produce two optical images on the focal plane. The relative brightness of the two split images can be adjusted by the degree of tangential motion of the P1 panels. 
Therefore, this pair of S1 panels can be aligned by applying rotation to one of the two S1 panels so that the pair of split images is merged (as illustrated by Figure~\ref{fig:S2_alignment_overview} (B)). 

The pSCT optical support structure (OSS) incorporates twice as many M1 panels as M2 panels. Therefore, the on-axis incident light that reaches each S1 panel is first reflected from two P1 panels. These two P1 panels can be rotated in opposite directions so that any misalignment between the S1 panel and both of its neighboring S1 panels are detected. Following this pattern, all 16 P1 panels are rotated about their radial axes in alternating directions, creating a maximum of 32 images if all S1-S1 edges are misaligned. Figure~\ref{fig:S2_alignment_overview} (C) shows an example of such an OSS configuration, with the zoom-in inset illustrating the split image pairs due to S1-S1 misalignment. 

Applying a tip or tilt motion of an S1 panel will cause the separation of two pairs of split images to change (due to the chosen alternating direction of the applied P1 rotation), and a response matrix can be measured. 
With the response matrices of S1 panels as well as the separation between the split images, S1 panels can be aligned with respect to each other. 
Note that not all images have a split counterpart, because three S1-S1 edges are blocked by the OSS support beams. For these three S1-S1 edges, pointing at a $1.2$ degrees offset from a bright star, in combination with an appropriate degree of P1 rotation, has been verified to successfully avoid the beam obscuration. As of June 2021, the tip/tilt of all S1 panels were aligned using the method described above.

%%%%%%%%%%%%%%%%%%%%%%%%%%%%%%%%%%%%%%%%%%%%%%%%%
% Elevation dependence
% Deivid + Ruo
%%%%%%%%%%%%%%%%%%%%%%%%%%%%%%%%%%%%%%%%%%%%%%%%%
\section{Elevation Dependence of On-Axis PSF}\label{sec:elevation_psf}
% - used 0th order correction to get to panels to defocused image aligned
% - Review 3 step process, takes about 30mins. (1. measure 3 rings + correct, 2. collapse S2 and correct 3. collapse to focus) - is this the sequence?
% - 3 arcmin maintained. Caveat: measurement time vs implementation has few mins delay
% - Can say there is prelim study on single panel PSF.

The structural deformations due to changing gravitational load of the pSCT OSS at different elevation angles leads to deviations in panel alignment and affects the PSF in non-negligible ways. In this section, we describe how we directly measure the on-axis optical PSF as a function of elevation to quantify the effect of the structural deformations. 

The process of aligning the panels to create a focused star on the focal plane assumes a single set of rotations of all panels from the same initial state. In this initial state, the panels are positioned in the predetermined "ideal" configuration for the de-focused star pattern. To align the mirror panels to fully focused configuration at  any elevation, the same set of tip/tilt rotations should be applied if the panels are positioned in the predetermined ideal de-focused star pattern in that elevation. A database was created to save this starting state for all panel positions for each elevation.

The procedure to create the database consisted of three steps, designed to maintain internal alignment between panels for each ring pattern and a consistent central position. In the first step, the centroids for the P1 ring were found on the focal plane and corrections were found for the ideal ring configuration. After moving these panels to the ideal P1 ring configuration, the focal plane image was again digitized to find both P1 and P2 ring centroids. This process was repeated a third time to correct the S2 panel positions (see inset in (B) of Figure \ref{fig:optical_alignment_overview}). With all three rings corrected, the state was saved for this elevation angle and the entire procedure was repeated for another elevation angle.

All of the panels were then moved using the default set of rotations to collapse the rings into the focal point (see inset in (A) of Figure~ \ref{fig:optical_alignment_overview}). 
The database of the de-focused star patterns per elevation was used to interpolate the actuator lengths of all panels at any given elevation. Once interpolated, this state was loaded to reproduce the de-focused star pattern, and then the default rotations were applied to align the system.

\begin{figure}[h]
    \centering
    \includegraphics[width=.7\textwidth]{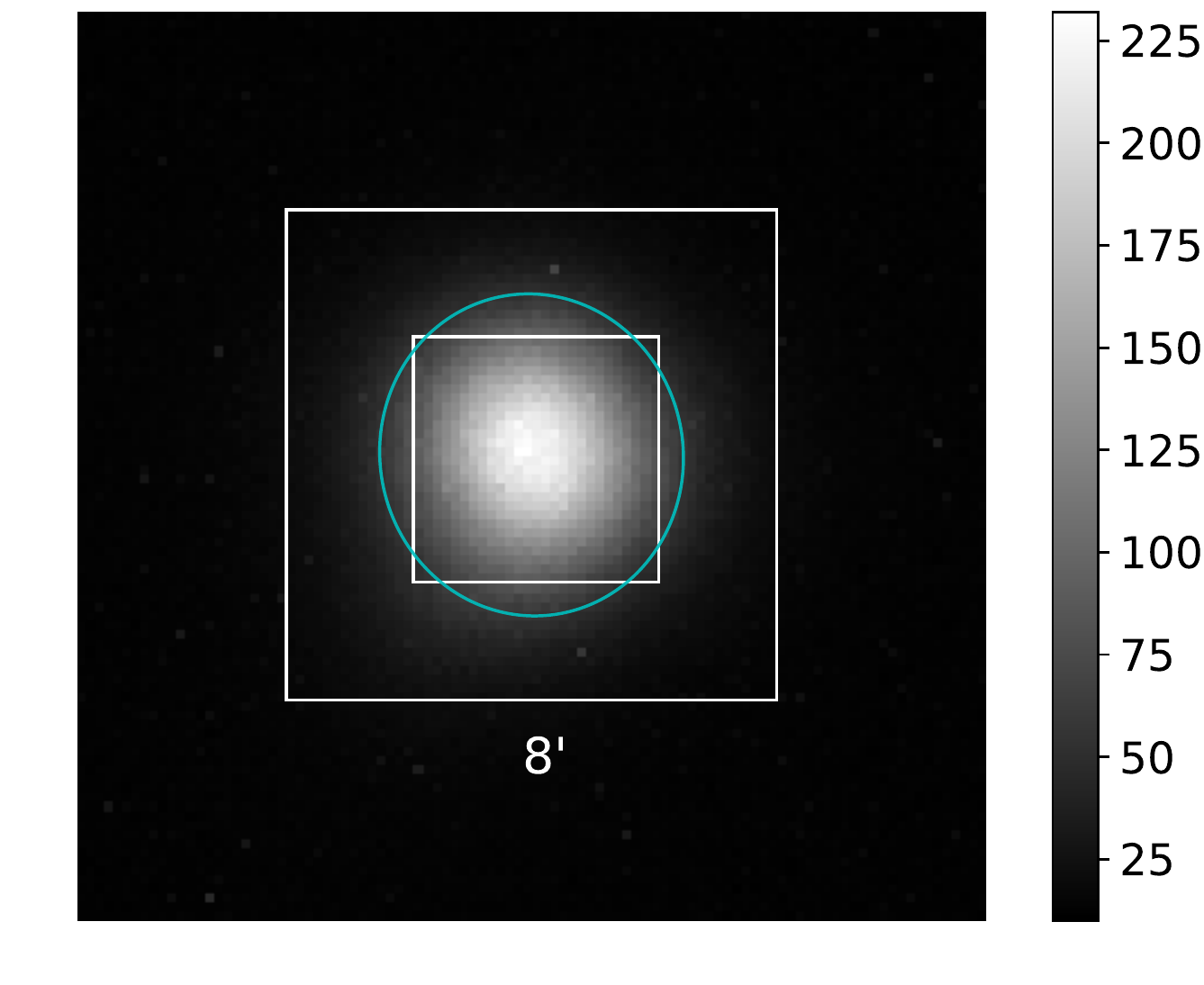}
    \caption{Image of Arcturus at 76\textdegree~ elevation. The PSF is 2.93'. The boxes indicate the relative size of the imaging (8') and trigger (4') pixels of the SiPM camera. }
    \label{fig:PSF_at_76el}
\end{figure}

The resulting on-axis PSF in the focused configuration was measured at each elevation. The PSF is defined as $2\times\textrm{max}(\sigma_x, \sigma_y)$ from the fit of a 2-D Gaussian on the centroid image. This value is converted to arcmin using the plate scale of the CCD pointing at the focal plane (0.241/1.625 arcmin/px). An example of the focal point is shown in Figure~\ref{fig:PSF_at_76el}, showing Arcturus at 76\textdegree~ elevation with a PSF of $2.93$'.

\begin{figure}[t]
    \centering
    \includegraphics[width=0.6\textwidth]{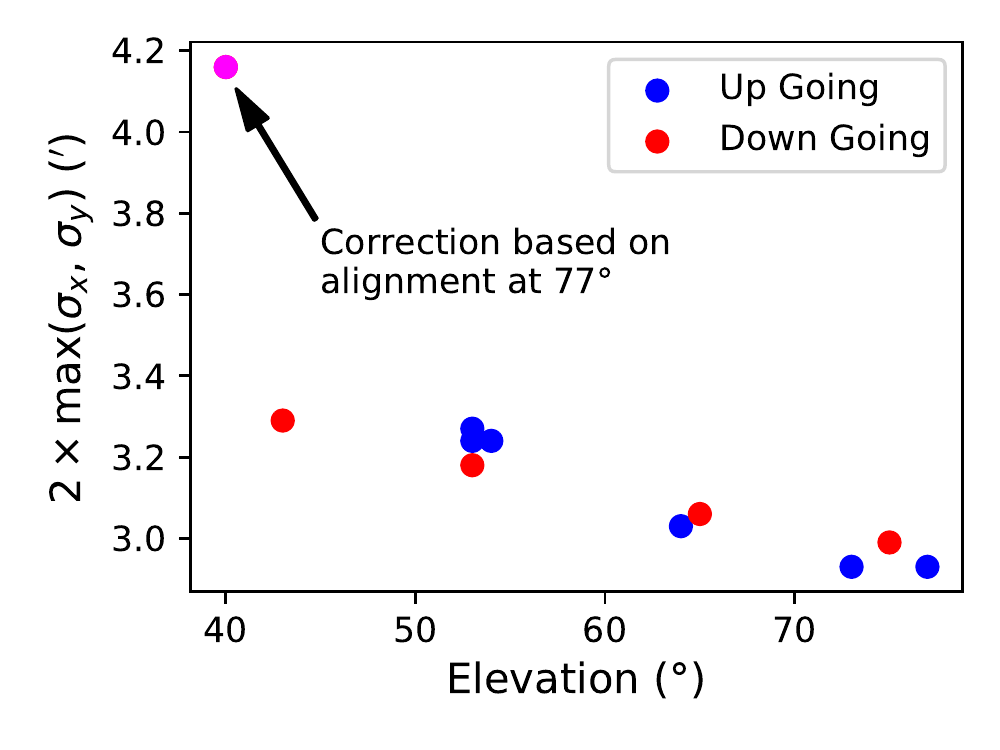}
    \caption{PSF measured per elevation. PSF is defined as $2\times \textrm{max}(\sigma_x, \sigma_y)$. At 40\textdegree, the panel alignment setting from 77\textdegree~ was loaded to measure the largest PSF deviation from the optimal setting, shown in purple. Panel alignment settings derived without corrections for all elevations.}
    \label{fig:elevation_psf}
\end{figure}

Figure \ref{fig:elevation_psf} shows the PSF in arcmin as a function of elevation achieved using outlined above methodology. The overall behavior is a stable PSF averaging 3.1' and getting better with increased elevation. To illustrate the necessity of elevation corrections of PSF, we aligned the pSCT OS at 77\textdegree~ and performed observations at 40\textdegree~ elevation which resulted in significantly degraded PSF of $4.16$'. This larger PSF is highlighted in Figure~\ref{fig:elevation_psf} to illustrate how large the PSF can get if active re-alignment is not attempted. This difference shows the negative impact of failing to align with the proper elevation-dependent state, revealing the effect of the structural deformation between extreme elevation angles on the PSF. 

In the process of measuring the elevation dependent alignment states, response matrices mapping panel rotations to 2-D centroid translations were used to position the panels in the focus position without non-linear correction, i.e. using the zeroth-order response matrices describing linear approximation. Iterating the motions of individual panels to minimize the offset from the target focus positions constitute finding the first-order corrections, which is part of an ongoing campaign. Applying the first-order corrections will improve the net PSF on all corrected panels since each panel centroid will be co-centered with every other panel, minimizing the spread of centroid offsets. Initial results indicate that it is possible to achieve alignment of the P1 ring in which the PSF of the ring is entirely dominated by the PSF of individual panels and the spreads of centroids is suppressed to a negligible contribution. 

The stability of the PSF based on pointing is an encouraging sign that full alignment corrections may be applied to the telescope while operating. The linear interpolation of actuator lengths between measured elevation states is a successful step that can be applied for any elevation. The speed of panel motion coupled with the relatively minor structural deformation dependent corrections (<0.5 mm in actuators) indicate that alignment may be kept stable throughout an observation, either actively at some frequency or in between observation runs.

%%%%%%%%%%%%%%%%%%%%%%%%%%%%%%%%%%%%%%%%%%%%%%%%%
% Conclusions + next steps
% 
%%%%%%%%%%%%%%%%%%%%%%%%%%%%%%%%%%%%%%%%%%%%%%%%%
\section{Summary and Outlook}\label{sec:summary}

In these proceedings, a method was developed to improve the alignment of the secondary mirror and the optical PSF was measured for a range of elevation angles. The method to align S1 was proved successful. This is a crucial step that ensures the correct tip/tilt alignment of all mirror panels of the pSCT and off-axis pSCT performance. The optical PSF was stable for a wide range of elevation angles, validating the creation of a database of panel configurations that can be used for active alignment corrections based on elevation.

% zeroth-order motions gave 3.1 which is worse than ray tracing estimates for ideal alignment 
% This preliminary measure is also uncorrected, and will be subject to improvement after applying first order corrections.

Ray-tracing studies indicated an ideal on-axis optical PSF better than the zeroth-order states measured per elevation angles here, which were stable at 3.1'. Future work on the optical system includes measurement of the first-order corrections and PSF per panel. Single panel PSF will provide the lower bound for the optical PSF alignment range, so each panel PSF will be studied. Additionally, off-axis alignment will be measured and verified and the GAS components will be commissioned. 

% goal: develop methods to maintain PSF (~3) across wide range of operating conditions. Previous work did P1, P2, S2. This update show further alignment + development of corrections based on pointing
% conclusion: we did it, here's how well we're doing and what remains: off-axis alignment, GAS, single panel PSF and P1 alignment -> further work on P2. 

%%%%%%%%%%%%%%%%%%%%%%%%%%%%%%%%%%%%%%%%%%%%%%%%%
%  Contribution notes for writing
%%%%%%%%%%%%%%%%%%%%%%%%%%%%%%%%%%%%%%%%%%%%%%%%%
% Contribution
% - Ruo: Pattern alignment
% - Qi: S1 alignment
% - Deivid: software to do this and the general pages. Wrangling
% - Steve F: Simulation pictures for S1 alignment. Goal was to understand the shadowing, but these pictures can be put together to show simulation v focal plane images.

%%%%%%%%%%%%%%%%%%%%%%%%%%%%%%%%%%%%%%%%%%%%%%%%%
% References
%%%%%%%%%%%%%%%%%%%%%%%%%%%%%%%%%%%%%%%%%%%%%%%%%
\bibliographystyle{ICRC}
\bibliography{references_ICRC}
% \printbibliography[heading=bibintoc] 

%%%%%%%%%%%%%%%%%%%%%%%%%%%%%%%%%%%%%%%%%%%%%%%%%
% Full authors list (ONLY FOR COLLABORATIONS) %%%%%%%%%%%%%%%%%%%%%%%%%%%%%%%%%%%%%%%%%%%%%%%%%
\clearpage
\section*{Full Authors List: \Coll\ Consortium}

\scriptsize
\noindent
C.~B.~Adams$^{1}$, 
G.~Ambrosi$^{2}$, 
M.~Ambrosio$^{3}$, 
C.~Aramo$^{3}$, 
P.~I.~Batista$^{4}$, 
W.~Benbow$^{5}$, 
B.~Bertucci$^{2, 6}$, 
E.~Bissaldi$^{7, 8}$, 
M.~Bitossi$^{9}$, 
A.~Boiano$^{3}$, 
C.~Bonavolont\`a$^{3, 10}$, 
R.~Bose$^{11}$, 
A.~Brill$^{1}$, 
J.~H.~Buckley$^{11}$, 
R.~A.~Cameron$^{12}$, 
R.~Canestrari$^{13}$, 
M.~Capasso$^{14}$, 
M.~Caprai$^{2}$, 
C.~E.~Covault$^{15}$, 
D.~Depaoli$^{16, 17}$, 
L.~Di~Venere$^{7, 8}$, 
M.~Errando$^{11}$, 
S.~Fegan$^{18}$, 
Q.~Feng$^{14}$, 
E.~Fiandrini$^{2, 19}$, 
A.~Furniss$^{20}$, 
A.~Gent$^{21}$, 
N.~Giglietto$^{7, 8}$, 
F.~Giordano$^{7, 8}$, 
E.~Giro$^{22}$, 
R.~Halliday$^{23}$, 
O.~Hervet$^{24}$, 
T.~B.~Humensky$^{25}$, 
S.~Incardona$^{26, 27}$, 
M.~Ionica$^{2}$, 
W.~Jin$^{28}$, 
D.~Kieda$^{29}$, 
F.~Licciulli$^{8}$, 
S.~Loporchio$^{7, 8}$, 
G.~Marsella$^{26, 27}$, 
V.~Masone$^{3}$, 
K.~Meagher$^{21, 30}$, 
T.~Meures$^{30}$, 
B.~A.~W.~Mode$^{30}$, 
S.~A.~I.~Mognet$^{31}$, 
R.~Mukherjee$^{14}$, 
D.~Nieto$^{32, 33}$, 
A.~Okumura$^{34}$, 
N.~Otte$^{21}$, 
F.~R.~Pantaleo$^{7, 8}$, 
R.~Paoletti$^{35, 9}$, 
G.~Pareschi$^{36}$, 
F.~Di~Pierro$^{17}$, 
E.~Pueschel$^{4}$, 
D.~Ribeiro$^{1}$, 
L.~Riitano$^{30}$, 
E.~Roache$^{5}$, 
J.~Rousselle$^{37}$, 
A.~Rugliancich$^{9}$, 
M.~Santander$^{28}$, 
R.~Shang$^{38}$, 
L.~Stiaccini$^{35, 9}$, 
L.~P.~Taylor$^{30}$, 
L.~Tosti$^{2}$, 
G.~Tovmassian$^{39}$, 
G.~Tripodo$^{26, 27}$, 
V.~Vagelli$^{40, 2}$, 
M.~Valentino$^{10, 3}$, 
J.~Vandenbroucke$^{30}$, 
V.~V.~Vassiliev$^{38}$, 
D.~A.~Williams$^{24}$, 
P.~Yu$^{38}$, 
\\ 

\noindent
$^{1}$Physics Department, Columbia University, New York, NY 10027, USA. 
$^{2}$INFN Sezione di Perugia, 06123 Perugia, Italy. 
$^{3}$INFN Sezione di Napoli, 80126 Napoli, Italy. 
$^{4}$Deutsches Elektronen-Synchrotron, Platanenallee 6, 15738 Zeuthen, Germany. 
$^{5}$Center for Astrophysics | Harvard \& Smithsonian, Cambridge, MA 02138, USA. 
$^{6}$Dipartimento di Fisica e Geologia dell’Universit\`a degli Studi di Perugia, 06123 Perugia, Italy. 
$^{7}$Dipartimento Interateneo di Fisica dell’Universit\`a e del Politecnico di Bari, 70126 Bari, Italy. 
$^{8}$INFN Sezione di Bari, 70125 Bari, Italy. 
$^{9}$INFN Sezione di Pisa, 56127 Pisa, Italy. 
$^{10}$CNR-ISASI, 80078 Pozzuoli, Italy. 
$^{11}$Department of Physics, Washington University, St. Louis, MO 63130, USA. 
$^{12}$Kavli Institute for Particle Astrophysics and Cosmology, SLAC National Accelerator Laboratory, Stanford University, Stanford, CA 94025, USA. 
$^{13}$INAF IASF Palermo, 90146 Palermo, Italy. 
$^{14}$Department of Physics and Astronomy, Barnard College, Columbia University, NY 10027, USA. 
$^{15}$Department of Physics, Case Western Reserve University, Cleveland, Ohio 44106, USA. 
$^{16}$Dipartimento di Fisica dell’Universit\`a degli Studi di Torino, 10125 Torino, Italy. 
$^{17}$INFN Sezione di Torino, 10125 Torino, Italy. 
$^{18}$LLR/Ecole Polytechnique, Route de Saclay, 91128 Palaiseau Cedex, France. 
$^{19}$Y. 
$^{20}$Department of Physics, California State University - East Bay, Hayward, CA 94542, USA. 
$^{21}$School of Physics \& Center for Relativistic Astrophysics, Georgia Institute of Technology, Atlanta, GA 30332-0430, USA. 
$^{22}$INAF Osservatorio Astronomico di Padova, 35122 Padova, Italy. 
$^{23}$Dept. of Physics and Astronomy, Michigan State University, East Lansing, MI 48824, USA. 
$^{24}$Santa Cruz Institute for Particle Physics and Department of Physics, University of California, Santa Cruz, CA 95064, USA. 
$^{25}$Science Department, SUNY Maritime College, Throggs Neck, NY 10465. 
$^{26}$Dipartimento di Fisica e Chimica "E. Segr\`e", Universit\`a degli Studi di Palermo, via delle Scienze, 90128 Palermo, Italy. 
$^{27}$INFN Sezione di Catania, 95123 Catania, Italy. 
$^{28}$Department of Physics and Astronomy, University of Alabama, Tuscaloosa, AL 35487, USA. 
$^{29}$Department of Physics and Astronomy, University of Utah, Salt Lake City, UT 84112, USA. 
$^{30}$Department of Physics and Wisconsin IceCube Particle Astrophysics Center, University of Wisconsin, Madison, WI 53706, USA. 
$^{31}$Pennsylvania State University, University Park, PA 16802, USA. 
$^{32}$Institute of Particle and Cosmos Physics (IPARCOS), Universidad Complutense de Madrid, E-28040 Madrid, Spain . 
$^{33}$Department of EMFTEL, Universidad Complutense de Madrid, E-28040 Madrid, Spain. 
$^{34}$Institute for Space--Earth Environmental Research and Kobayashi--Maskawa Institute for the Origin of Particles and the Universe, Nagoya University, Nagoya 464-8601, Japan. 
$^{35}$Dipartimento di Scienze Fisiche, della Terra e dell'Ambiente, Universit\`a degli Studi di Siena, 53100 Siena, Italy. 
$^{36}$INAF - Osservatorio Astronomico di Brera, 20121 Milano/Merate, Italy. 
$^{37}$Subaru Telescope NAOJ, Hilo HI 96720, USA. 
$^{38}$Department of Physics and Astronomy, University of California, Los Angeles, CA 90095, USA. 
$^{39}$Instituto de Astronom\'ia, Universidad Nacional Aut\'onoma de M\'exico, Ciudad de M\'exico, Mexico. 
$^{40}$Agenzia Spaziale Italiana, 00133 Roma, Italy.

\end{document}

%% file: ICRC2021 template/optical_alignment_overview.tex
The optical PSF is most sensitive to the tip/tilt rotation of the M1 and M2 panels and also the translations of the M2 panels. These degrees of freedom are constrained using de-focused images of stars.
This section provides a brief summary of the optical alignment procedure when the inner ring of S1 panels of M2 is assumed to be aligned (see next section). For more detailed information on this alignment procedure and earlier results, see \cite{Adams2020, Adams2020a}.

The de-focused image of a star is created by the reflection from a pair of M1 and M2 panels in which each panel is purposely misaligned to form patterns of rings of individual pairs so that the 80 images created by all possible pairs in the OS are identifiable.  

Images of a star produced by pairs of panels P1-S1, P2-S1, and P2-S2 are arranged into three concentric ring configurations, which are shown in Figure~\ref{fig:optical_alignment_overview}: the aligned panels in the focused configuration are shown in Figure~\ref{fig:optical_alignment_overview} (A), while two different states of the de-focused configuration are shown in Figures~\ref{fig:optical_alignment_overview} (B) and (C).
These de-focused configurations facilitate easy identification of images produced by each combination of primary and secondary panels and enables simultaneous and efficient alignment of each image into the predetermined pattern, which must be produced by a star. In addition, the characteristics of individual images (major and minor axes and elongation) are used to guide the relative global positioning of M1, M2 and FP.

To relate the motion of the images on the FP to the adjustments of the actuators of Stewart platforms, the response matrices for each panel are simultaneously measured at the 3-ring configuration (Figure~\ref{fig:optical_alignment_overview} (C)) by rotating (tip and tilt) each mirror panel and observing the motion of the corresponding focal-plane image.

Based on the response matrices, the displacement between predetermined pattern of pairs of images of the star with respect to the focal point can be used to guide the precise tip/tilt rotation of each panel in order to achieve the focus of the OS. This procedure using measured response matrices constitutes the zeroth-order alignment of the mirror panels, which is not yet ideal because the response matrices are measured at the 3-ring configurations instead of at the focused configuration where linear approximation may be insufficient. To further improve alignment, the development of the 1st-order alignment correction at the focal point is currently in progress.

% \cite{Adams2020} % Qi proceeding, SPIE
% \cite{Adams2020a} % Ruo proceeding, SPIE
% \cite{Rousselle13} %spie
% \cite{Nieto15} % icrc15
% \cite{Rousselle2015} % spie
% \cite{Nieto:2017cyj} % icrc17

%% file: main.bbl
\providecommand{\href}[2]{#2}\begingroup\raggedright\begin{thebibliography}{1}

\bibitem{leslie}
L.~P. Taylor, {\it Design and performance of the prototype schwarzschild-couder
  telescope camera},  in {\em these proceedings}, PoS (ICRC2021) 748, 2021.

\bibitem{Adams2020}
C.~Adams and {others}, {\it {Verification of the optical system of the 9.7-m
  prototype Schwarzschild-Couder Telescope}},  in {\em Optical System
  Alignment, Tolerancing, and Verification XIII} (J.~Sasián and R.~N.
  Youngworth, eds.), vol.~11488, pp.~10 -- 28, International Society for Optics
  and Photonics, SPIE, 2020.

\bibitem{Adams2020a}
C.~B. Adams and {others}, {\it {Alignment of the optical system of the 9.7-m
  prototype Schwarzchild-Coulder Telescope}},  in {\em Ground-based and
  Airborne Telescopes VIII} (H.~K. Marshall, J.~Spyromilio, and T.~Usuda,
  eds.), p.~96, SPIE, Dec., 2020.

\bibitem{Vassiliev07}
V.~{Vassiliev}, S.~{Fegan}, and P.~{Brousseau}, {\em Astroparticle Physics}
  {\bf 28} (Sept., 2007) 10--27.

\bibitem{2015arXiv150902345O}
A.~N. {Otte} and {others}, {\em arXiv e-prints} (Sept., 2015) arXiv:1509.02345.

\bibitem{Sreenivasan94}
S.~{Sreenivasan}, K.~{Waldron}, and P.~{Nanua}, {\em Mechanism and Machine
  Theory} {\bf 29} (1994) 855 -- 864.

\bibitem{mahnke2009opc}
W.~Mahnke and S.-H. Leitner, {\em ABB Review} {\bf 3} (2009) 3.

\end{thebibliography}\endgroup


\providecommand{\href}[2]{#2}\begingroup\raggedright\endgroup
